\documentclass[conference]{IEEEtran}
\IEEEoverridecommandlockouts
\usepackage{amsmath}
\usepackage{graphicx}
\usepackage{amssymb}
\usepackage{cases}
\usepackage{cite}
\usepackage[ruled,vlined,lined,ruled,linesnumbered]{algorithm2e}
\usepackage{relsize}
\usepackage[nodisplayskipstretch]{setspace}
\usepackage{fancyhdr}
\begin{document}
\title{Wireless Power Charging Control in Multiuser Broadband Networks}

\author{Suzhi~Bi and Rui Zhang \\
        ECE Department, National University of Singapore, Singapore \\ E-mail:\{bsz,~elezhang\}@nus.edu.sg \vspace{-2ex}}
\maketitle

\fancyhead[CO,CE]{\small{This paper has been accepted by \emph{IEEE ICC Workshop} 2015.}}

\vspace{-1.8cm}
\begin{abstract}
Recent advances in wireless power transfer (WPT) technology provide a cost-effective solution to charge wireless devices remotely without disruption to the use. In this paper, we propose an efficient wireless charging control method for exploiting the frequency diversity in multiuser broadband wireless networks, to reduce energy outage and keep the system operating in an efficient and sustainable state. In particular, we first analyze the impact of charging control method to the operating lifetime of a WPT-enabled broadband system. Based on the analysis, we then propose a multi-criteria charging control policy that optimizes the transmit power allocation over frequency by jointly considering the channel state information (CSI) and the battery state information (BSI) of wireless devices. For practical implementation, the proposed scheme is realized by a novel limited CSI estimation mechanism embedded with partial BSI, which significantly reduces the energy cost of CSI and BSI feedback. Simulation results show that the proposed method could significantly increase the network lifetime under stringent transmit power constraint. Reciprocally, it also consumes lower transmit power to achieve near-perpetual network operation than other single-criterion based charging control methods.
\end{abstract}

\section{Introduction}
The performance of wireless systems is fundamentally constrained by limited device battery life. As frequent battery replacement/recharging by manual operation is often costly and inconvenient, numerous energy-conservation methods have been proposed to prolong the operating lifetime of wireless communication networks, via transmit power control, energy-aware medium access control (MAC) and routing selection, etc \cite{2002:Biyikoglu,2004:Younis,2007:Chen}. Alternatively, radio frequency enabled (RF-enabled) \emph{wireless power transfer} (WPT) technologies provide an attractive solution to power wireless devices (WDs) over the air \cite{2014:Bi}. By leveraging the far-field radiative properties of microwave, WDs could harvest energy remotely from the RF signals radiated by the dedicated energy transmitter \cite{2013:Zhou}. Currently, RF power in milliwatt (mW) level could be effectively transferred to WDs from a distance of more than $10$ meters.\footnote{See Powercast Corp. website at http://www.powercastco.com.} The received RF energy is sufficient to power the activities of many low-power devices, such as sensors and RF identification (RFID) tags, where some commercial RF-enabled WPT products have already been developed by Powercast Corp. Besides, the recent development of MIMO technology significantly boosts the energy transfer efficiency \cite{2013:Zhang,2014:Xu}, which also opens up more potential applications for RF-enabled WPT in the future.

The practical advantages of WPT have recently attracted increasing efforts to develop wireless powered communication systems \cite{2014:Bi,2013:Zhou,2013:Zhang,2014:Ju1,2014:Zhou,2013:Nintanavongsa,2014:Xu}. Among them, WPT for broadband networks is of particular interest, where both wireless information and power transfer could benefit from the broadband channel diversity \cite{2014:Zhou,2013:Nintanavongsa}. However, unlike information transfer, WPT does not introduce detrimental interference to unintended co-channel receivers. This fundamental difference motivates this study to redesign the \emph{MAC mechanism} in WPT-enabled broadband systems for efficient charging coordination among the distributed WDs, which are of general power usage besides communication purpose.

Efficient WPT control in broadband systems requires the energy transmitter to have accurate knowledge of both the channel state information (CSI) and battery state information (BSI) of WDs. While accurate CSI could effectively enhance the energy transfer efficiency, well-informed BSI helps reduce energy outage rate by transmitting on the sub-channels in favor of those close-to-outage WDs. Although being assumed by many previous studies for the simplicity of analysis (e.g., \cite{2014:Zhou,2013:Nintanavongsa}), in practice, however, achieving perfect knowledge of CSI/BSI consumes significant amount of energy for the WDs to perform CSI estimation and CSI/BSI feedback. The energy cost to the WDs may even offset the power gain obtained from more refined transmit power allocation.

In this paper, we propose a new cost-effective charging control method in WPT-based broadband networks. The proposed method has the following main features and advantages:
\begin{enumerate}
  \item A multi-criteria charging policy is proposed based on the analysis of expected network lifetime of WPT-based systems. Specifically, different metrics are used to evaluate the energy-harvesting performance of the WDs depending on their residual battery levels or BSI.
  \item The charging control policy is implemented by a novel limited CSI estimation mechanism embedded with partial BSI, which incurs very low complexity and energy cost on CSI/BSI feedback. In particular, simulations show that good system performance is achievable even with very limited CSI feedback, e.g., by estimating at most $4$ out of $50$ sub-channels for a WD to feed back.
  \item The network lifetime performance of the proposed multi-criteria method is \emph{robust} against different network deployment, and could be further optimized by tuning the control parameters according to specific applications.
\end{enumerate}
The performance advantage of the proposed multi-criteria charging control method is evaluated through extensive simulations and comparisons with other heuristically designed single-criterion based benchmark schemes.

\section{System Model}
We consider a WPT-enabled broadband wireless network, where an energy node (EN) is connected to a stable power source and broadcasts RF energy to power $K$ distributed WDs. Each WD has an RF energy harvesting circuit and rechargeable battery of capacity $C$ to store the harvested energy and power the activities of the WD, e.g., event sensing or data transmission. The bandwidth of the system is $W$ Hz, which is equally divided into $N$ sub-channels each with bandwidth $W/N$. The wireless channels are assumed to be reciprocal in the uplink and downlink, and independent among the WDs. For the simplicity of exposition, we assume that the channel experiences block fading, where the sub-channel gains remain constant in a transmission block of length $T$ and vary independently over different blocks. Notice that the block fading assumption does not affect the validity of our analysis or the operation of the proposed protocols.

In the $j$-th transmission block, $j\in \{0,1,\cdots\}$, we use $h_{k,i}^{j}$ to denote the channel power gain of the $i$-th sub-channel between the EN and the $k$-th WD, $P_i^{j}$ to denote the transmit power of the EN on the $i$-th sub-channel. The total transmit power is bounded by $\sum_{i=1}^N P_i^{j}\leq P_0$, $j=0,1,\cdots$. Then, the harvested energy by the $k$-th WD within the block is \cite{2013:Zhou}
\begin{equation}
\small
\label{1}
Q_k^{j} = \eta T \sum_{i=1}^N P_i^{j} h^{j}_{k,i},\ \ k=1,\cdots, K,
\end{equation}
where $\eta \in (0,1]$ is a fixed parameter denoting the energy harvesting efficiency which is assumed to be known by the EN. Let $X_k^{j}$ denote the residual energy of the $k$-th WD at the end of $j$-th transmission block, and $E_k^{j}$ denote the amount of energy consumed within the transmission block. Within each block, we assume that the energy consumption rate is constant so that the energy level increases or decreases monotonically. Then, the residual energy at the end of the $(j+1)$-th block is
\begin{equation}
\small
\label{2}
X_k^{j+1} = \min\left\{\max\left(X_k^{j}-E_k^{j+1} + Q_k^{j+1},0\right),C\right\}.
\end{equation}
In this paper, $E_k^{j}$ is assumed to follow a general distribution with an average consumption rate $\mathbb{E}[E_k^{j}]= \mu_k T$, $\forall j$.

The output voltage of a battery decreases with the residual energy level. We say an \emph{energy outage} occurs if the remaining energy of a WD $k$ is no larger than a certain threshold $\nu^l$, such that normal device operation could not be maintained. Without loss of generality, we set $\nu^l=0$ throughout this paper. Given the initial battery level $\mathbf{X}^{0}=\left[X_1^0,\cdots,X_K^0\right]$, we define \emph{network lifetime} as the duration until a WD is in energy outage. Whenever an energy outage occurs, the device in energy outage will enter a hibernation mode, which harvests RF energy and returns to normal operation only after the battery level reaches a prescribed threshold $\nu^u$ ($\nu^u>\nu^l$). This is a common practice to avoid frequent energy outage for continuous device operation. Notice that the results obtained from homogeneous battery assumption here could also be extended to heterogeneous case (i.e., WDs have different capacities and threshold parameters) by scaling the respective channel gains and energy consumption rates accordingly.

Due to channel diversities among the WDs over both time and frequency, the network lifetime is determined by the transmit power allocation strategy $\mathbf{P}^j=[P_1^j,\cdots,P_N^j]$ over time $j=1,2,\cdots$. Besides, the knowledge of CSI and BSI, although costly to obtain, is critical to the performance of transmit power allocation. In the following, we first analyze the impact of charging control method to the network lifetime. Based on the analysis, we then propose an efficient charging control method with limited CSI/BSI feedback mechanism.

\section{Charging Control Policy Analysis}
A point to notice is that the network lifetime is affected by many factors besides transmit power allocation, such as EN placement, total transmit power and channel distributions. In this section, we assume that all the other parameters are fixed to focus on the design of transmit power allocation policy. We denote $L_\psi$ as the expected network lifetime achieved by a charging policy $\psi$. Specifically, a charging control policy $\psi$ specifies the transmit power allocation in each transmission block, and thus the harvested energy $Q_k^{j}$, for $k=1,\cdots,K$ and $j=1,2,\cdots$. To void trivial results, we assume that the expected network lifetime is finite regardless of the charging policy used, i.e.,
\begin{equation}
\label{7}
\small
\underset{\psi \in \pi}{\text{maximize  }}L_\psi < \infty,
\end{equation}
where $\pi$ is the set of all feasible policies that satisfy the transmit sum-power constraint, $P_0$. In other words, $P_0$ is assumed to be sufficiently small, so that the total energy harvesting rate is always lower than the consumption rate, to be consistent with the condition in (\ref{7}). In fact, as verified later by simulations, a charging policy $\psi$ that achieves a longer $L_\psi$ under a given transmit power budget also requires lower transmit power to achieve near-perpetual network operation (i.e., $L_\psi$ is a very large number). Therefore, finding the optimal charging policy under finite network lifetime assumption also has important implication in practical system designs with the guarantee of sufficiently large $L_\psi$.

\subsection{Wireless Charging: A Game-theoretic View}
The actual battery dynamic in (\ref{2}) complicates the analysis because of the max/min operators, yet unable to provide extra insight into charging policy design. To make the problem tractable, we adjust the battery model as follows:
\begin{enumerate}
  \item The energy level could surpass the battery capacity at the end of a transmission block, but such an overcharged battery cannot harvest any energy in the following transmission block(s) until the energy level drops below the capacity due to power usage;
  \item When an energy outage occurs, the residual energy could be negative at the end of the current transmission block.
\end{enumerate}
The first modification helps remove the min operator in (\ref{2}), while ensuring that the energy level is approximately upper bounded by the capacity $C$. Specifically, it overestimates the harvested energy when battery is fully-charged in a transmission block, but underestimates the harvested energy in the following transmission block(s). The second modification removes the max operator in (\ref{2}). Both modifications have limited effect to the modeling accuracy, because a practical transmission block length is sufficiently small, such that the energy harvested/consumed within a block is marginal compared to the battery capacity. With the two minor modifications, we could express the battery dynamics as a random process $X_k^{j+1} = X_{k}^{j} - E_{k}^{j+1} + \bar{Q}_{k}^{j+1}, \ \ j=0,1,\cdots$, where
\begin{equation}
\small
\bar{Q}_{k}^{j+1}= \begin{cases}
0, & X_{k}^{j}\geq C,\\
Q_{k}^{j+1}, & \text{otherwise}
\end{cases}
\end{equation}
denotes the harvested energy in the $(j+1)$-th transmission block of the modified battery model, with $Q_{k}^j$ given in (\ref{1}). The average energy harvesting rate is determined by the control policy $\psi$, and denoted by $\lambda_k^{\psi} \triangleq \mathbb{E}[Q_{k}^{j}]/T$. Accordingly, an energy outage occurs if $X_{k}^{j}\leq 0$.

Equivalently, the charging problem could be modeled by a gambling game with the $K$ WDs as gamblers. $X_k^{j}$ is the balance of gambler $k$, who continuously plays a betting game with $\bar{Q}_{k}^{j+1}$ as the income and $E_{k}^{j+1}$ as the loss in the $(j+1)$-th bet, $j=0,1,\cdots$. The game starts with each gambler holding $X_k^0$ initial balance, and stops once a gambler's balance becomes zero or negative. Then, the stopping time of the game is also the network lifetime. Evidently, the game is unfair because the average income and loss of each bet are not equal in general, i.e., $\lambda_k^{\psi} \neq \mu_k$. In particular, a gambler $k$ receives zero income in the $(j+1)$-th bet when $X_k^{j}\geq C$. In the following, we derive the expected stopping time of the above gambling game by constructing an auxiliary fair game, such that Martingale stopping time theorem \cite{2001:Grimmett} could be applied.

\subsection{Expected Network Lifetime}
We first introduce an auxiliary random process $\mathbf{Z}_j = [Z_1^{j},Z_2^{j},\cdots, Z_K^{j}]$, $j=0,1,\cdots$, where $Z_{k}^j \triangleq X_{k}^j + Y_{k}^j$ and
\begin{equation*}
\small
Y_k^j= \begin{cases}
0, & j=0,\\
Y_k^{j-1} + \left(\mu_k-\lambda_k^{\psi}\right)T, & j>0 \text{ and } X_{k}^{j-1} < C,  \\
Y_k^{j-1} + \mu_k T,  &   j>0 \text{ and } X_{k}^{j-1} \geq C.
\end{cases}
\end{equation*}
$Y_k^j$ could be considered as the cumulative compensation given to gambler $k$ up to the $j$-th bet, where the gambler is compensated for $\left(\mu_k-\lambda_k^{\psi}\right)T$ in the $j$-th bet if its balance before the bet, i.e., $X_{k}^{j-1}$, is less than $C$. Otherwise, it receives $\mu_k T$ if $X_{k}^{j-1} \geq C$. The following lemma shows that the random process $\mathbf{Z}_j$ is a Martingale.

\emph{Lemma 1:} The random process $\left\{\mathbf{Z}_j,j\geq 0\right\}$ is a Martingale, or equivalently the bet is a fair game.

\emph{Proof:} To prove Lemma $1$, we need to show that for all $j$ it satisfies 1) $\mathbb{E}\left[\mathbf{Z}_j\right] <\infty$ and 2) $\mathbb{E}\left[\mathbf{Z}_{j+1} | \mathbf{Z}_j = \mathbf{z}_j, \cdots, \mathbf{Z}_1 = \mathbf{z}_1\right] = \mathbf{z}_j$ \cite{2001:Grimmett}. Condition $1$) holds from the assumption that $L_{\psi}$ (the number of bets) is finite. For condition $2$), we have for each $k$
\begin{equation*}
\small
\begin{aligned}
&\mathbb{E}\left[Z_{k}^{j+1}\big|Z_k^j = z_k^j, \cdots, Z_k^0 = z_k^0\right] \\
= &z_k^j + \mathbb{E}\left[\bar{Q}_{k}^{j} - \mathbb{E}_{k}^{j}\right] + \mathbf{1}_{kj}^{C}\cdot \mu_k T + \left(1- \mathbf{1}_{ij}^{C}\right)\cdot(\mu_k-\lambda_k^{\psi})T = z_k^j,
\end{aligned}
\end{equation*}
where $\mathbf{1}_{kj}^{C}$ is an indicator function that equals $1$ if $X_k^j\geq C$ and $0$ otherwise. This completes the proof.  $\hfill \blacksquare$

Then, the following Martingale Stopping Theorem \cite{2001:Grimmett} could be used to derive the expected network lifetime.

\emph{Proposition 1 (Martingale Stopping Theorem):} Let $\left\{\mathbf{Z}_j,j\geq 0\right\}$ be a Martingale and $W$ a stopping time that depends only on the value of $\mathbf{Z}_j$. If $\mathbb{E}\left[|\mathbf{Z}_W|\right]<\infty$, then $\mathbb{E}\left[\mathbf{Z}_W\right]=\mathbb{E}\left[\mathbf{Z}_0\right]$.

In our case, $W$ corresponds to the number of bets until $X_k^W\leq 0$ for some $Z_k^W$. For an initial state $\mathbf{Z}_0$, we have
\begin{equation}
\label{8}
\small
\mathbb{E}\left[\sum_{k=1}^K Z_k^0\right] = \sum_{k=1}^K \left(\mathbb{E}\left[X_k^0\right] + \mathbb{E}\left[Y_k^0\right]\right) = \sum_{k=1}^K x_k^0 \triangleq \varepsilon_0,
\end{equation}
where $x_k^0$ is the value of $X_k^0$, and $\varepsilon_0$ denotes the initial sum-energy of the WDs. Similarly, at the end of the $W$-th bet, we have
\begin{equation}
\label{9}
\small
\begin{aligned}
&\mathbb{E}\left[\sum_{k=1}^K Z_k^W\right] = \mathbb{E}\left[\sum_{k=1}^K X_k^W\right] + \mathbb{E}\left[\sum_{k=1}^K Y_k^W\right] \\
 & \triangleq \varepsilon_r^{\psi} + \mathbb{E}\left[W\right]T \left[ \sum_{k=1}^K \mu_k - \sum_{k=1}^K\left(1-\alpha_k^{\psi}\right)\lambda_k^{\psi}\right] ,
\end{aligned}
\end{equation}
where $\alpha_k^{\psi}$ denotes the probability that the battery of the $k$-th WD is fully- or overcharged, and $\varepsilon_r^{\psi} \triangleq \mathbb{E}\left[\sum_{k=1}^K X_k^W\right]$ denotes the expected residual sum-energy achieved by policy $\psi$. From Proposition $1$, we could infer that the righthand sides of (\ref{8}) and (\ref{9}) are equal. With simple calculations, the expected stopping time conditioned on $\varepsilon_0$ is obtained as
\begin{equation}
\small
\label{4}
\mathbb{E}\left[L_{\psi}|\varepsilon_0\right]=\mathbb{E}\left[W\right]T = \frac{\varepsilon_0 - \varepsilon_r^{\psi}}{\sum_{k=1}^K \mu_k  - \sum_{k=1}^K(1-\alpha_k^{\psi})\lambda_k^{\psi}}.
\end{equation}
We notice that $E\left[L_{\psi}|\varepsilon_0\right]$ is always positive by our assumption that the total energy harvesting rate is smaller than the consumption rate, i.e., $\sum_{k=1}^K \lambda_k^{\psi} \leq \sum_{k=1}^K \mu_k$, $\forall \psi \in \pi$. Besides, the network lifetime expression in \cite{2007:Chen} for conventional battery-powered wireless sensor network is a special case of ours when $\lambda_k^{\psi}=0$, i.e., WPT is not used.

\subsection{Charging Policy Analysis}
To prolong the expected network lifetime in (\ref{4}), a charging policy $\psi$ should produce
\begin{enumerate}
  \item low total residual energy $\varepsilon_r^{\psi}$;
  \item high total energy harvesting rate $\sum_{k=1}^K \lambda_k^{\psi}$;
  \item and low overcharging probabilities $\alpha_k^{\psi}$'s.
\end{enumerate}
For condition $1$), the ideal scenario is for all the WDs to have zero residual energy simultaneously. In this case, the optimal charging policy is the \emph{max-min} approach, i.e., maximizing the minimum residual energy among the WDs after each energy transmission block. For the second condition, the optimal policy is the \emph{max-rate} approach, i.e., maximizing the total harvested energy by all the WDs in each transmission block. For the third condition, the optimal policy is the \emph{min-max} approach, i.e., minimizing the maximum residual energy among the WDs after each transmission block. In a practical wireless powered network, however, the above three charging policies may conflict with each other. For instance, the max-rate approach is strongly biased towards nearby WDs with high average channel gains. The power allocation of the max-min approach, however, is in favor of WDs that are far-away from the EN with smaller channel gains. This simple fairness-efficiency tradeoff indicates that the three objectives could not be satisfied by a single-criterion charging policy in general.

To balance the three objectives, we therefore consider a multi-criteria approach. Our observation is that overcharging only occurs to close-to-capacity WDs. Thus, the EN should apply a min-max criterion to the close-to-capacity WDs, to discourage allocating power to the sub-channels that may lead to overcharge in the coming transmission block(s). Besides, using max-min approach to charge the currently close-to-outage WDs could effectively reduce $\varepsilon_r^{\psi}$, because otherwise larger amount of energy will be harvested by the WDs of moderate/high energy levels in the subsequent transmission block(s), leading to higher $\varepsilon_r^{\psi}$. For those WDs of moderate energy levels, the max-rate approach is most suitable, because charging them with the highest efficiency does not have immediate effect to either $\alpha_k^{\psi}$'s or $\varepsilon_r^{\psi}$, but could effectively enhance the sum charging rate $\sum_{i=k}^K \lambda_k^{\psi}$.

The above analysis motivates a multi-criteria charging control method that jointly considers the CSI and BSI. However, a practical question arises on how to efficiently feed back CSI/BSI to the EN in an energy-constrained broadband system with a large number of WDs and sub-channels. To tackle this problem, we propose in the following a simple yet efficient charging control protocol based on limited CSI/BSI feedback.

\section{Proposed Protocol}

\begin{figure}
\centering
  \begin{center}
    \includegraphics[width=0.37 \textwidth]{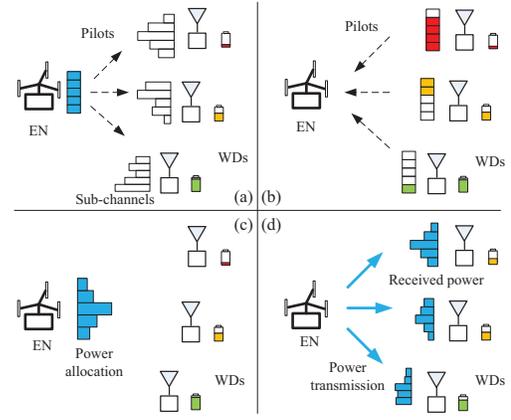}
  \end{center}
  \caption{Illustration of the charging control protocol. Sub-figures (a)-(d) correspond to steps $1)$-$4)$ of the protocol descriptions, respectively. }
  \label{11}
\end{figure}

In this section, we propose a multi-criteria charging control protocol using a limited CSI feedback mechanism. The protocol operates in the following steps and is illustrated in Fig.~$\ref{11}$. Without causing confusions, we omit the transmission block index $j$ in this section for the brevity of notation.
\begin{enumerate}
  \item At the beginning of a transmission block, the EN sends pilot signals on the $N$ sub-channels. Then, each WD $k$ estimates its own sub-channel power gains, i.e., $h_{k,i}$'s;
  \item Depending on the residual energy $X_k$, each WD $k$ sends back to the EN orthogonal narrowband pilot signals on its $M_k$ strongest sub-channels, with
      \begin{equation}
      \label{10}
      \small
M_k = \begin{cases}
m_l, & X_k\leq \tau_l,\\
m_m, & \tau_l < X_k \leq \tau_h,  \\
m_h,  &   X_k > \tau_h,
\end{cases}
\end{equation}
where $m_h\leq m_m \leq m_l$ are positive integers, and $0 < \tau_l,\tau_h< C$ are two predetermined energy thresholds. The subset of sub-channels reported by WD $k$ is denoted by $\mathcal{M}_k$, $k=1,\cdots,K$.
\item Here, we assume that the channel estimation by the EN is perfect. Besides the knowledge of the channel gains, i.e., $\left\{h_{k,i}|i\in \mathcal{M}_k, i=1,\cdots,N, k=1,\cdots,K \right\}$, the EN is also aware of the quantized BSI of each WD by counting the number of pilots sent by the WDs, i.e., $M_k$'s. The EN then optimizes its transmit power allocation on all the sub-channels using a policy detailed later in (\ref{5}).
\item The WDs harvest RF energy in the remaining transmission block. Then, the iteration repeats from Step $1)$.
\end{enumerate}
Instead of sending pilots on all the sub-channels, each WD only reports on a small subset of sub-channels with the largest channel gains, which significantly reduces the energy cost. The intuition behind is that transmit power is in general only allocated to relatively strong sub-channels, thus the knowledge of weak sub-channels' CSI has marginal effect to the power allocation solution. Although consuming more energy to send pilots on a larger number of sub-channels, from (\ref{10}), a WD of lower residual energy could in fact benefit from a more favorable power allocation due to the better knowledge of CSI by the EN. On the other hand, the number of pilots sent by a WD also contains partial BSI indicating the energy level, which will be exploited by our proposed power control. In addition, the values of $\left\{m_l,m_m,m_h\right\}$ could be tuned in different scenarios, e.g., setting $m_m = m_h=0$ such that a WD sends energy request only when it is close to outage. The impact of these parameters will be shown by simulations.

The knowledge of CSI and BSI enables a multi-criteria charging policy as discussed in Section III.C. In particular, from the values of $M_k$'s, the EN is aware of the subsets of WDs in low, moderate and high residual energy levels, denoted by $\mathcal{N}_l,\mathcal{N}_m$ and $\mathcal{N}_h$, respectively. We design the EN to allocate transmit power $P_i$ to the $i$-th sub-channel, $i=1,\cdots,N$, by solving the following linear programming (LP) problem:
\begin{equation}
\small
\label{5}
   \begin{aligned}
    & \underset{\mathbf{P}}{\text{maximize}} & &  \underset{k\in \mathcal{N}_l}{\text{min}} \hat{Q}_k + c_1 \mathsmaller\sum_{k\in \mathcal{N}_m} \hat{Q}_k - c_2 \underset{k\in \mathcal{N}_h}{\text{max}} \hat{Q}_k\\
    & \text{subject to}    & & \mathsmaller\sum_{i=1}^N P_i = P_0,\ \ \ P_i \geq 0, \ i=1,\cdots,N,
   \end{aligned}
\end{equation}
where $\hat{Q}_k$ denotes the estimated harvested energy by the $k$-th WD, given by $\hat{Q}_k = \eta T \cdot \mathsmaller\sum_{i=1}^N P_i \hat{h}_{k,i}$, where
\begin{equation}
\small
\label{6}
\hat{h}_{k,i} = \begin{cases}
h_{k,i}, & i\in \mathcal{M}_k,\\
\mathbb{E}\left[h_{k,i}\big|h_{k,i}\leq \text{min}_{j\in \mathcal{M}_k} h_{k,j}\right] , & i\notin  \mathcal{M}_k.
\end{cases}
\end{equation}
The three terms in the objective of (\ref{5}) denote the weighted energy harvested by WDs in $\mathcal{N}_l,\mathcal{N}_m$ and $\mathcal{N}_h$, respectively. $c_1$ and $c_2$ are two positive weights set for balancing the three terms. Specifically, the first term maximizes the minimum harvested energy of the close-to-outage WDs. The second term maximizes the total harvested energy by the WDs with moderate energy level. The last term discourages the EN to charge the close-to-capacity WDs. Generally speaking, a larger (smaller) $c_1$ would enhance the energy harvesting efficiency (fairness). A larger $c_2$ would increase the penalty to charge the WDs that are close-to-capacity. In general, $c_1$ and $c_2$ are set much smaller than $1$ to ensure that priority is given to the close-to-outage WDs.

Note that from (\ref{6}), $\hat{Q}_k$ is calculated using both the \emph{explicit} CSI reported on $\mathcal{M}_k$, and the \emph{implicit} CSI for those sub-channels not reported by WD $k$. In (\ref{6}), we use conditional expectation to calculate $\hat{h}_{k,i}$ given that the channel gains of unreported sub-channels are lower than the known channel gains in $\mathcal{M}_k$. Then, the calculation of $\hat{h}_{k,i}$ for $i\notin  \mathcal{M}_k$ is determined by the distribution of $h_{k,i}$'s. Denote $t_k \triangleq \text{min}_{i\in \mathcal{M}_k} h_{k,i}$. For instance, the estimate is $\hat{h}_{k,i} = h^l + t_k/2$ if $h_{k,i}$'s are independent and uniformly distributed within $[h^l,h^h]$. If $h_{k,i}$'s follow independent exponential distribution with mean $h_0$, the estimate is $\hat{h}_{k,i} = t_k + h_0 - t_k/\left(1 - e^{-t_k/h_0}\right)$.

The proposed charging control protocol incurs low implementation cost. All the computations are borne by the EN, while each energy-constrained WD only needs to send out limited number of pilots based on the measurement of channel gains and its own residual energy level.

\section{Performance Evaluation}
In this section, we evaluate the performance of the proposed multi-criteria charging (MCC) control algorithm. Unless otherwise stated, the parameters used in all simulations are listed in Table I, which correspond to a typical indoor sensor/RFID network. Without loss of generality, the path loss exponent is $2$, such that the path loss is roughly $37.7$ dB at $2$ meters from the EN. In practice, multiple ENs are needed to cover a large area, which is beyond the scope of this paper and will be considered in our future work. The wireless channel power gains follow exponential distributions with mean obtained from the path loss model. Here, we consider an i.i.d. energy consumption model where a WD consumes $28$ mW power with probability $0.25$ within a block, and no power with probability $0.75$. In this case, the average energy consumption rate is $7$ mW, where a fully-charged battery will be depleted in about $2.5$ hours without WPT. The initial battery level of all WDs is $80\%$ of the capacity, and $\eta=1$ is assumed.

\begin{table}
\caption{Simulation Parameters}
\footnotesize
\begin{center}
\begin{tabular}{|c|c||c|c|}
 \hline
  EN Tx power &   $1$ W & Path loss exponent &   $2$\\ \hline
  Center frequency &   $915$ MHz &  Tx block length &   $100$ ms \\ \hline
  No. of SCs &   $50$  & Ave. WD power cons.&   $7$ mW\\ \hline
  SC bandwidth &   $10$ KHz   & Battery voltage &   $3$ V\\ \hline
  Tx antenna gain &   $2.5$  & Battery capacity &   $6$ mAh\\ \hline
  Rx antenna gain &   $2$  & Pilot Tx power per SC &   $0.02$ mW\\ \hline
\end{tabular}
\end{center}
\label{stat}
\end{table}

For the proposed MCC algorithm, we set $\left\{\tau_l,\tau_h\right\}= \left\{2.4,5.4\right\}$ mAh and $\left\{m_l,m_m,m_h\right\}=\left\{4,2,1\right\}$. That is, a WD with less than $40\%$ of the battery capacity will report the $4$ best sub-channels, while those with more than $90\%$ battery will report $1$ sub-channel. Otherwise, a WD will report on $2$ sub-channels. Besides, we set $c_1=0.2$ and $c_2=0.1$ in (\ref{5}). We have also considered three single-criterion based benchmark schemes for comparison, including
\begin{enumerate}
  \item UNI: uniform power allocation, i.e., $P_i = P_0/N$, $\forall i$.
  \item MaxRate:  maximize the total harvested energy by the WDs, i.e., $\text{max} \sum_{k=1}^{K} \hat{Q}_k$.
  \item MaxMin: maximize the minimum harvested energy by the WDs of the lowest energy level. That is, $\text{max} \text{ min}_{k\in \mathcal{N}_l} \hat{Q}_k$ when $\mathcal{N}_l \neq \emptyset$, or $\text{max} \text{ min}_{k\in \mathcal{N}_m} \hat{Q}_k$ when $\mathcal{N}_l=\emptyset$ and $\mathcal{N}_m \neq \emptyset$, otherwise $\text{max} \text{ min}_{k\in \mathcal{N}_h} \hat{Q}_k$ when $\mathcal{N}_l=\mathcal{N}_m=\emptyset$, where $\emptyset$ denotes empty set.
\end{enumerate}
The case of MinMax method (minimize the maximum energy harvested) is omitted here due to its poor charging performance when it is used alone. Notice that the UNI scheme is completely oblivious to the CSI and BSI, thus the WDs do not need to send any pilot. The MaxRate utilizes only CSI and the MaxMin approach uses both CSI and BSI. For fair comparisons, these two methods use the same CSI feedback method as the MCC algorithm.

\begin{figure}
\centering
  \begin{center}
    \includegraphics[width=2.9in]{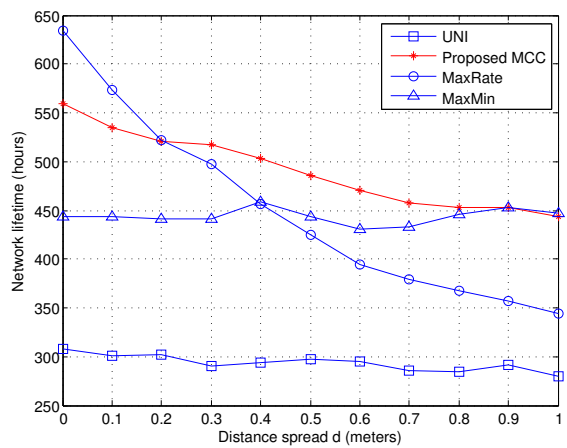}
  \end{center}
  \caption{Comparisons of average network lifetime vs. distance spread $d$.}
  \label{12}
\end{figure}

In Fig.~$\ref{12}$, we compare the average network lifetime achieved by the $4$ methods. We assume that $6$ WDs are uniformly placed within a ring region $\left[2-d/2,2 + d/2\right]$ meters with the EN being the center, where a smaller (larger) $d$ leads to lower (higher) distance disparity to the EN. Fig.~$\ref{12}$ shows the lifetime of different schemes as $d$ increases. Each point in the figure is the average performance of $10$ random placements, while the lifetime of each placement is the mean of $10$ independent simulations. We see that UNI has the worst performance for not being able to exploit the channel diversity. We also see that charging efficiency dominates the lifetime when $d$ is small, where MaxRate performs the best when $d\leq 0.2$. However, its performance degrades drastically as $d$ increases, as its charging control is strongly biased towards the nearby WDs. On the other hand, user fairness dominates when $d$ is large, where the MaxMin approach achieves the longest network lifetime for $d\geq 0.9$. In between, the proposed MCC method achieves the best performance for $0.2\leq d \leq 0.9$. Overall, it achieves over $10\%$ longer lifetime than MaxRate and MaxMin, and over $65\%$ performance gain than the UNI method. More importantly, it is robust in different placement scenarios, unlike the MaxRate and MaxMin methods that perform poorly when $d$ is either large or small. In addition, the performance of MCC could be further optimized according to the specific placement by adjusting the parameters $c_1$ and $c_2$. We observe in experiments that a smaller $c_1$ is preferred when the network size is large and the WDs are more sparsely deployed, i.e., large $d$. Detailed results are omitted here due to the page limit.

\begin{figure}
\centering
  \begin{center}
    \includegraphics[width=2.9in]{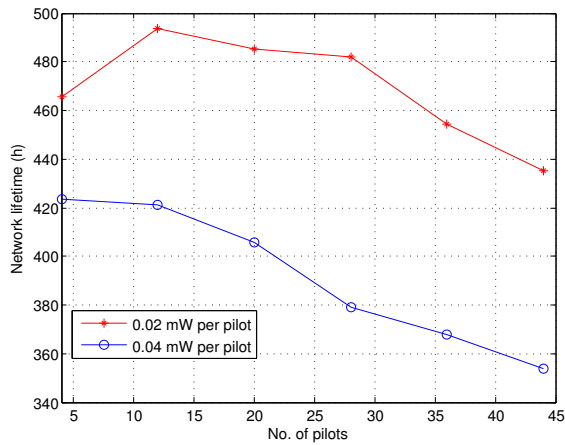}
  \end{center}
  \caption{Network lifetime of MCC vs. no. of CSI feedbacks ($d=0.6$ m).}
  \label{13}
\end{figure}

In Fig.~$\ref{13}$, we plot the average network lifetime as the amount of feedback varies. Here, we fix $d$ as $0.6$m. The x-axis is the value of $m_l$. For convenience, we set $m_m = m_l/2$ and $m_h=m_l/4$. We see that when the transmit power per sub-channel (SC) is $0.02$ mW,  $m_l =12$ achieves the longest network lifetime. However, as $m_l$ increases, energy consumed on CSI feedback would eventually offset the energy gain from more refined power allocation, resulting in a decreased lifetime. This is more evident as the transmit power increases to $0.04$ mW per SC, where the maximum lifetime is achieved when the minimum value of $m_l$ allowed in this case is attained, i.e., $m_l=4$. The results show that for the proposed MCC, the pilot energy consumption plays a critical role in the resulting network lifetime.

\begin{figure}
\centering
  \begin{center}
    \includegraphics[width=2.9in]{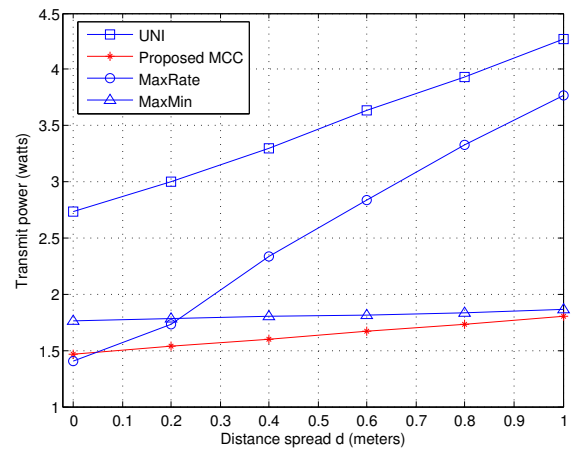}
  \end{center}
  \caption{Minimum transmit power to achieve perpetual network operation.}
  \label{15}
\end{figure}

We also plot in Fig.~$\ref{15}$ the minimum power required by different charging methods to achieve near-perpetual network operation. Due to the randomness of channel fading and energy consumptions, it is not possible to truly sustain perpetual network operation in practice. Here, we say a WPT-enabled system is near-perpetual if the network lifetime is longer than $3000$ hours in all the $10$ independent simulations conducted. For fair comparison, we deploy the $6$ WDs in a line equally spaced from $2-d/2$ to $2+d/2$. Not surprisingly, the UNI method requires the highest transmission power in all the placements. The transmit power of MaxRate is the lowest when $d=0$, but increases drastically as $d$ increases. In contrast, MCC and MinMax require much lower transmit power than the other two methods. The power increment is slow as the distance disparity among the WDs increases. In particular, the proposed MCC outperforms the MinMax method due to its higher energy transfer efficiency. The results in Figs.~\ref{12} and \ref{15} show that a scheme that achieves a longer network lifetime under a given transmission power constraint, is in general also more power-efficient to achieve self-sustainable operation. This also justifies the practical value of our analytical results in Section III under the finite network lifetime assumption.

\section{Conclusions}
In this paper, we have proposed a novel multi-criteria charging control method to prolong the lifetime of a wireless powered multiuser broadband system. The proposed charging policy is efficiently implemented using a limited CSI feedback mechanism embedded with partial BSI. Simulations results show that the proposed charging control could significantly increase the network lifetime under a fixed transmit power constraint, and is also power efficient to achieve self-sustainable network operations.

\end{document}